\documentclass[3p,times,procedia]{elsarticle}
\usepackage{ecrc2}
\usepackage{textcomp}
\usepackage[bookmarks=false]{hyperref}
    \hypersetup{colorlinks,
      linkcolor=blue,
      citecolor=blue,
      urlcolor=blue}

\usepackage{amssymb}

\usepackage[figuresright]{rotating}
\usepackage{graphicx}
\usepackage{dcolumn}
\usepackage{bm}
\usepackage{xcolor} 
\usepackage{amsmath}

\begin{document}
\begin{frontmatter}




\title{\textit{accLB}: A High-Performance Lattice Boltzmann Code for Multiphase Turbulence on Multi-Gpu Architectures }


\author[a]{Marco Lauricella} 
\author[b]{Aritra Mukherjee}
\author[b,c]{Luca Brandt}
\author[a,d,e]{Sauro Succi} 
\author[f]{Daulet Izbassarov}
\author[g]{Andrea Montessori \corref{cor1}}

\address[a]{Istituto per le Applicazioni del Calcolo, Consiglio Nazionale delle Ricerche, Via dei Taurini 19, Rome, 00185, Italy}
\address[b]{Department of Energy and Process Engineering, Norwegian University of Science and Technology (NTNU), Trondheim, Norway}
\address[c]{Department of Environment, Land and Infrastructure Engineering (DIATI), Politecnico di Torino, Torino, Italy}
\address[d]{Center for Life Nano- \& Neuro-Science, Fondazione Istituto Italiano di Tecnologia, Viale Regina Elena 295, Rome, 00161, Italy}
\address[e]{Department of Physics, Harvard University, 17 Oxford St., Cambridge, 02138, MA, USA}
\address[f]{Finnish Meteorological Institute, Erik Palmenin aukio 1, Helsinki, 00560, Finland}
\address[g]{Department of Civil, Computer Science and Aeronautical Technologies Engineering, Roma Tre University, Via Vito Volterra, Rome, 00146, Italy}
\begin{abstract}

In this work, we present accLB, a high-performance Fortran-based lattice Boltzmann (LB) solver tailored to multiphase turbulent flows on multi-GPU architectures. The code couples a conservative phase-field formulation of the Allen–Cahn equation with a thread-safe, regularized LB method to capture complex interface dynamics. Designed from the ground up for HPC environments, accLB employs MPI for distributed memory parallelism and OpenACC for GPU acceleration, achieving excellent portability and scalability on leading pre-exascale systems such as Leonardo and LUMI. Benchmark tests demonstrate strong and weak scaling efficiencies on multiple GPUs. Physical validation includes direct numerical simulations of homogeneous isotropic turbulence (HIT). Further, we examine bubble-laden HIT and observe a transition to a $-3$ energy scaling, as in experiments and theoretical predictions, due to bubble-induced dissipation, along with enhanced small-scale intermittency. These results highlight accLB as a robust and scalable platform for the simulation of multiphase turbulence in extreme computational regimes.

\end{abstract}

\end{frontmatter}

\email{andrea.montessori@uniroma3.it}

\vspace*{-6pt}

\section{Introduction}\label{intro}
Multiphase turbulent flows play a central role in a wide array of natural phenomena and industrial processes, from cloud formation and ocean--atmosphere interactions to chemical reactors and spray combustion~\cite{jahne1998, schramm2003, grabowski2013growth, sardina2015}. Accurately modeling these flows is particularly challenging due to the complex interplay between turbulence and dispersed phases, which span a vast range of spatial and temporal scales~\cite{brandt2022ARFM, crialesi2023interaction}. Even single-phase turbulence exhibits multiscale dynamics, from the large energy-containing eddies down to the small-scale dissipative motions at the Kolmogorov scale. In multiphase systems, the complexity deepens as interface phenomena push the smallest relevant scales down to the molecular level.

Capturing these coupled multiscale features with high fidelity requires direct numerical simulations (DNS), which are notoriously computationally demanding---especially when resolving deformable fluid interfaces. As such, the development of scalable numerical solvers tailored for high-performance computing (HPC) platforms is essential to advance our understanding of turbulent multiphase flows.

The advent of exascale computing has opened unprecedented opportunities for tackling such grand-challenge simulations. According to the 64th \textit{Top500} list, the leading exascale systems---El Capitan, Frontier, and Aurora---are hosted in the United States, while Europe is home to four pre-exascale machines among the global top ten: HPC6 (Italy), Alps (Switzerland), LUMI (Finland), and Leonardo (Italy). These machines are built on heterogeneous architectures, combining powerful CPUs with GPU accelerators such as AMD GPUs (HPC6 and LUMI) or NVIDIA GPUs (Alps and Leonardo), pushing the boundaries of performance and energy efficiency.

Modern HPC systems, particularly those leveraging GPU-based architectures, have significantly expanded the scope of direct numerical simulations in computational fluid dynamics (CFD). Several open-source CFD solvers, including AFiD~\cite{Zhu_AFiD_2018}, STREAmS~\cite{Bernardini2021}, CaNS~\cite{costa2021}, TLBfind \cite{pelusi2022tlbfind,pelusi2023analysis},  and LBcuda~\cite{Bonaccorso2022}, have been optimized for NVIDIA GPUs using CUDA Fortran. However, maintaining performance portability across different GPU vendors remains a pressing challenge. To overcome this, recent efforts have embraced directive-based programming models, such as OpenACC, which offer a path toward cross-platform compatibility. Codes like FluTAS~\cite{crialesi2023flutas}, CaNS-Fizzy~\cite{lupo2025}, and the pseudo-spectral DNS solver ported by \citet{Roccon2024} exemplify this approach. Most recently, URANOS-2.0~\cite{devanna2025} showcased a portable OpenACC-based implementation for both AMD and NVIDIA platforms.

In this work, we present accLB, a high-performance, thread-safe lattice Boltzmann solver specifically developed to simulate multiphase turbulent flows on modern GPU-accelerated supercomputers. The solver combines a conservative phase-field formulation based on the Allen--Cahn equation with a regularized LBM scheme, enabling accurate tracking of complex interface dynamics under large density and viscosity contrasts. Its architecture employs hybrid MPI--OpenACC parallelism, ensuring both scalability and portability across heterogeneous computing environments.

We validate the solver’s performance and accuracy through extensive benchmarks on EuroHPC pre-exascale systems Leonardo and LUMI, demonstrating both strong and weak scaling up to 64 GPUs and achieving sustained throughputs exceeding 150~GLUPS(Giga Lattice Updates Per Second). Direct numerical simulations of single- and two-phase homogeneous isotropic turbulence confirm the robustness and fidelity of the accLB framework \cite{lauricella2025thread} in capturing the rich dynamics of multiphase turbulence. 


\section{Methods}\label{method}

\subsection{Navier-Stokes equations for multiphase flows with interfaces}

The multiphase system analyzed in this study is described by the continuity and momentum equations, which are numerically solved using the lattice Boltzmann method (LBM). These equations read:
\begin{equation}
    \partial_t \rho + \partial_\alpha (\rho u_\alpha) = 0,
\end{equation}
\begin{equation} \label{NSeq}
     \partial_t (\rho u_\alpha) +  \partial_\beta( \rho u_\alpha u_\beta) = -\partial_\alpha p + \partial_\beta \left(\rho \nu [\partial_\beta u_\alpha + \partial_\alpha u_\beta]\right) + F^s_\alpha,
\end{equation}
where \( u_\alpha \) denotes the velocity field, \( \rho \) the fluid density, \( p \) the macroscopic pressure, \( \nu \) the kinematic viscosity, and \( F^s_\alpha \) the surface tension force. Greek indices represent Cartesian tensor components.

The surface tension force \( F^s_\alpha \) is expressed as the divergence of the capillary stress tensor, $F^s_\alpha = \partial_\beta \left(\gamma (\delta_{\alpha\beta} - n_\alpha n_\beta)\right)$,
where \( n_\alpha \) is the local interface normal and \( \gamma \) the surface tension coefficient.

\subsubsection{Allen-Cahn Equation for Interface Tracking}

The interface between the two fluid phases is tracked using a conservative phase-field formulation based on the Allen-Cahn equation. The scalar order parameter \( \phi \), which varies smoothly between 0 and 1 across the interface, evolves according to:

\begin{equation}
    \partial_t\phi + u_\alpha \partial_\alpha \phi = D \partial_\alpha\partial_\alpha \phi - \kappa \partial_\alpha (\phi(1-\phi)n_\alpha),
\end{equation}

where \( D \) denotes the interface diffusivity and \( \kappa = 4D/\delta \), with \( \delta \) representing the interface width. The interface is implicitly located at the iso-surface \( \phi = \phi_0 = 0.5 \). The equilibrium profile across an interface centered at \( x_0 \) is given by \cite{kim2005continuous},  $\phi(x, y, z)=\phi_0 \pm \frac{\phi_H - \phi_L}{2}\tanh\left( \frac{|x-x_0|+|y-y_0|+|z-z_0|}{\delta} \right)$.


This phase-field equation is tightly coupled to the hydrodynamic fields via the advection term. 

To accurately resolve interface dynamics in the presence of strong advection, the convective term in Eq. (4) is discretized using a fifth-order Weighted Essentially Non-Oscillatory (WENO-5) scheme \cite{JIANG1996202}. This high-order scheme minimizes numerical dissipation while capturing sharp gradients and reducing spurious oscillations, making it well-suited for turbulent multiphase flows.

\subsection{Thread-safe lattice Boltzmann model}

The LBM provides an efficient solver for the Navier-Stokes equations, particularly for complex or multiphase flows. In this work, we employ a thread-safe and regularized LBM framework, as described in \cite{montessori2023thread, lauricella2025thread}.

The second-order accurate discrete lattice Boltzmann equation over a set of $q$ velocity directions is given by \cite{succi,kruger2017lattice}:
\begin{equation} \label{lbe}
    f_i(x_\alpha+c_{i\alpha}\Delta t,t + \Delta t)=f_i(x_\alpha,t) + \omega(f^{eq}_{i}-f_i) + (1-\frac{\omega}{2})S_i,
\end{equation}
where $\omega$ is a relaxation rate, \( f_i \) are the distribution functions, \( f_i^{eq} \) their equilibrium counterparts, and \( S_i \) denotes an external forcing term incorporated through the Guo forcing scheme \cite{guo2002discrete}:
\begin{equation} \label{guoforce}
    S_i=w_i \left( \frac{c_{i\alpha} - u_\alpha}{c_s^2} + \frac{c_{i\beta} u_\beta}{c_s^4}c_{i\alpha}\right)F_\alpha,
\end{equation}
being \( F_\alpha \) the total body force acting on the fluid.

The relaxation time, $\tau=1/\omega$, controls the kinematic viscosity through the relation \( \nu = c_s^2(\tau - 0.5) \) \cite{succi,kruger2017lattice,montessori2018lattice}.

Defining $f^{neq}_{i}=f_i-f^{eq}_{i} + 0.5 S_i$ and substituting in Eq. \ref{lbe}, the post-collision update can be reformulated as \cite{feng2019hybrid,feng2019hybrid2}:
\begin{equation} \label{pushLB}
    f_i(x_\alpha+c_{i\alpha}\Delta t,t + \Delta t) = f^{eq}_{i} + (1-\omega)f^{neq}_{i} + \frac{1}{2}S_i.
\end{equation}

In the thread-safe implementation, both equilibrium and non-equilibrium components are reconstructed directly from macroscopic fields, avoiding direct access to full distribution functions and preventing race conditions due to non-local memory operations \cite{montessori2023thread,montessori2024high}. The equilibrium distribution reads, $f_i^{eq} = w_i \left( p^* + \frac{c_{i\alpha} u_\alpha}{c_s^2} + \frac{(c_{i\alpha}c_{i\beta} - c_s^2 \delta_{\alpha\beta}) u_\alpha u_\beta}{2 c_s^4} \right)$, 
where \( p^* = p / (\rho c_s^2) \) is the dimensionless pressure. The non-equilibrium contribution is reconstructed up to second order as $f_i^{neq}=\frac{c_{i\alpha}c_{i\beta} a^{(2)}_{1,\alpha\beta}}{2 c_s^4}$, 
where $a^{(2)}_{1,\alpha\beta}$ denotes the second order Hermite coefficient of non-equilibrium \cite{malaspinas2015increasing}.
Macroscopic variables are obtained from the distribution functions using $p^* (x_\alpha,t)=\sum_i f_i(x_\alpha,t)$ and $ u_{\alpha}(x_\alpha,t)=\sum_i f_i(x_\alpha,t)c_{i\alpha} + \frac{1}{2}\sum_i S_i$:
The second order Hermite coefficient, $a^{(2)}_{1,\alpha\beta}$, is assessed by the sum over the discrete non-equilibrium populations $f_i^{neq}$ weighted on the second order Hermite polynomial $  \mathcal{H}^{(2)}_{i,\alpha\beta} = c_{i,\alpha} c_{i,\beta} - c_s^2 \delta_{\alpha\beta}$ obtaining:
\begin{equation} \label{moments2}
    a^{(2)}_{1,\alpha\beta}(x_\alpha,t)=\sum_i f_i^{neq}(x_\alpha,t)(c_{i,\alpha} c_{i,\beta} - c_s^2 \delta_{\alpha\beta}).
\end{equation}
where, as shown in Ref. \cite{malaspinas2015increasing}, apart from higher-order smallness in the Mach number ($\mathrm{Ma}$), the second order coefficient macroscopically reads: $a^{(2)}_{1,\alpha\beta}=-c_s^2 \tau \Bigl[
  \partial_{\alpha}\Bigl(u_{\beta}\Bigr)\;+\;
  \partial_{\beta}\Bigl(u_{\alpha}\Bigr) \Bigr] + \mathcal{O}(\mathrm{Ma}^3)$.

\subsection{Extension for high-density and viscosity contrasts}

To accurately simulate multiphase flows with strong density and viscosity contrasts, additional force contributions are introduced into the momentum equation. The total force acting on the system is given by \cite{fakhari2017improved}:
\begin{equation}
    F_\alpha = F_\alpha^s + F_\alpha^p + F_\alpha^\nu,
\end{equation}
where \( F_\alpha^s \) is the surface tension force, computed via a phase-field-based formulation\cite{kim2005continuous}, $F_\alpha^s=-\gamma\kappa\nabla\phi|\nabla\phi|$.


The pressure-induced force reads, $F_\alpha^p = -p^* c_s^2 \partial_\alpha \rho$, 
where \( \rho = \phi \rho_l + (1 - \phi) \rho_g \) is the local mixture density. Given that the pressure is defined as \( p = p^* \rho c_s^2 \), its gradient contains implicit terms that require explicit correction via \( F_\alpha^p \).

Lastly, the viscous correction force is given by $F_\alpha^\nu = -\frac{\nu \omega}{c_s^2 \Delta t} \left[\sum_i (f_i - f_i^{eq}) c_{i\alpha} c_{i\beta} \right] \partial_\alpha \rho$:

Spatial derivatives appearing in the force terms and phase-field evolution equation are discretized using lattice-based stencils, $\partial_\alpha \Psi = \frac{1}{c_s^2} \sum_i w_i \Psi(x_\alpha + c_{i\alpha}) c_{i\alpha}$ and $\partial_\alpha \partial_\beta \Psi = \frac{1}{c_s^2} \left( \sum_{i \neq 0} w_i \Psi(x_\alpha + c_{i\alpha}) - w_0 \Psi(x_\alpha) \right)$

This modeling framework enables robust and efficient simulations of multiphase flows with large physical property mismatches while maintaining numerical stability and thread safety through the use of regularization techniques.

\section{Implementation}

The accLB code is developed in Fortran 95, utilizing modules to keep the code organized and reduce clutter. Variables related to specific features (such as fluids and phase-field) or methods (like time integrators) are grouped into separate modules for better structure. Portability and readability are key design goals of accLB, ensuring the code can run across multiple architectures and is accessible for contributors despite the complexity of the underlying model.

From its inception, accLB has been designed with parallel computing platforms in mind. It uses the Message Passing Interface (MPI), the standard for inter-node communication. However, a corresponding serial version of accLB is available and can be selected during compilation. 

The parallelization of the hydrodynamic solver in accLB is based on domain decomposition \cite{gropp1992parallel}, where the simulation domain is divided into one-, two-, or three-dimensional blocks distributed across MPI tasks. The block decomposition is selected at run-time, without the need to recompile the code, allowing users to optimize performance based on the global grid size and the number of available compute cores. For cubic domains, three-dimensional decomposition typically yields the best performance by minimizing the surface area involved in inter-process communication. The current implementation employs a halo region of two lattice spacings, which is necessary to calculate the derivative terms in the WENO5 scheme. This halo width is flexible and could be easily extended in future developments to accommodate more advanced algorithms.

The accLB code leverages OpenACC directives to enable execution on GPU devices, allowing for efficient offloading of compute-intensive tasks. This approach facilitates parallel acceleration with minimal code modifications, ensuring portability across different hardware architectures. By utilizing OpenACC, the code can fully utilize available GPU resources to enhance performance significantly, particularly for large-scale simulations. OpenACC also ensures broad compatibility across various GPU platforms, including devices from NVIDIA, AMD, and other vendors that support the standard. This flexibility allows the code to remain portable and adaptable to evolving hardware environments without requiring extensive rewrites.
Finally, OpenACC enables compilation and execution on purely CPU-based architectures as well. The code exploits a hybrid shared-distributed memory parallelization strategy in such cases, combining OpenACC and MPI to scale across multiple CPU cores and nodes efficiently. This unified parallel framework allows direct performance comparison across different high-performance computing platforms, such as CPU-based versus GPU-based clusters, aiding in performance benchmarking and resource optimization.

The time-marching implementation of the lattice-Boltzmann equation Eq. (\ref{lbe}) exploits the reformulation given in Eq. (\ref{pushLB}) to facilitate completely independent computations at each computational node, effectively eliminating race conditions. This design includes the thread-safe paradigm introduced by Montessori et al. \cite{montessori2025thread,montessori2024high,montessori2023thread}.
This approach enhances data locality and removes memory dependencies during the propagation step by reconstructing the post-collision distribution functions through Hermite projection. Such a strategy is particularly advantageous for parallel computing architectures as it ensures safe and efficient execution across multiple processing units without the need for complex synchronization mechanisms. Moreover, this method significantly reduces the memory footprint, enabling the simulation of large-scale fluid dynamics problems with improved performance and stability, achieving a balance between computational efficiency and the accuracy required for simulating complex fluid behaviors.
Thus, the primary variables in accLB include the 27 distribution functions alongside with the ten recovered macroscopic fields (dimensionless pressure $p*$, velocity vector $u_{\alpha}$, and six Hermite coefficient terms  $a^{(2)}_{1,\alpha\beta}$ of the corresponding symmetric tensor). The distribution functions can be stored using one of two competing memory layouts: array of structures (AoS) or structure of arrays (SoA), regardless of the underlying memory order (row-major as in C or column-major as in Fortran) \cite{exasc2}.
The accLB code employs the SoA layout, which has demonstrated superior performance over the AoS configuration. In particular, the SoA layout enables a unit-stride access pattern during the streaming step, ensuring coalesced memory access on GPU architectures.
This choice significantly improves memory efficiency, reducing bandwidth-related bottlenecks and making it the optimal strategy for high-performance implementations.

\section{Performances}

To evaluate performance, we simulate two different benchmark cases: 1) a single-component fluid and 2) a two-component system containing a single droplet of heavier density dispersed in a lighter continuous phase (density contrast equal to 100), both with a kinematic viscosity equal to 0.005 in lattice units. 

In the present work, we report both strong and weak scaling to assess the parallel performance of the code, where strong scaling provides information on how the execution time decreases with an increasing number of GPUs for a fixed problem size, reflecting the efficiency in accelerating a single workload. Instead, weak scaling analyzes how well the code maintains a constant execution time when the problem size grows proportionally with the number of GPUs, indicating scalability to larger workloads. The two scaling analyses provide a comprehensive view of parallel efficiency and the ability of accLB to handle increasingly demanding simulations.

Hence, the simulation domain is fixed as a cubic box with 512 lattice points per side in the strong scaling benchmarks. In the weak scaling simulations, we consider a cubic sub-domain of
side 512 lattice points for each GPU device (fixed sub-domain size), adopting a linear decomposition along the z-axis. Thus, we obtain an overall simulation box with sides 512, 512, and 512$*n_{p}$ with $n_{p}$ denoting the number of GPU devices for the weak scaling simulations.

Efficiency is assessed using the Giga Lattice Updates Per Second (GLUPS) metric, defined as: 
\begin{equation}
\label{eq:mlups}
\text{GLUPS}=\frac{L_x L_y L_z}{10^9 t_{\text{s}}},
\end{equation}
where $t_{\text{s}}$ is the run (wall-clock) time (in seconds) per single time step iteration, while $L_x$, $L_y$, and $L_z$ are the domain sizes.

In the present work, we define the speedup, $S_{p}$, in terms of GLUPS ratio as follows:
\begin{equation}
S_{p}=\frac{\text{GLUPS}_{p}}{\text{GLUPS}_{s}},
\label{eq:speedup}
\end{equation}
where $\text{GLUPS}_{s}$ denotes the Giga Lattice Updates Per Second for the code running on a single GPU and $\text{GLUPS}_{p}$ represents the Giga Lattice Updates Per Second for the code running on $n_{p}$ GPU cards. It is worth highlighting that the above definition of $S_{p}$ differs from the typical expression given in the literature \cite{hennessy2011computer}. Finally, the parallel efficiency, $E_{p}$, reads:
\begin{equation}
E_{p}=\frac{S_{p}}{n_{p}}.
\end{equation}

The benchmark simulations were conducted on two different GPU-based architectures: the pre-exascale Leonardo HPC system at CINECA in Italy, equipped with four NVIDIA Ampere\texttrademark~A100 Tensor Core GPU devices per computing node, each GPU card with 64 GB of HBM2e memory, and the LUMI pre-exascale supercomputer, hosted at CSC in Finland, 
based on the HPE Cray EX architecture, each compute node equipped with four AMD Instinct\texttrademark~MI250X GPUs, where each GPU device contains two Graphics Compute Dies (GCDs), with each die providing 64 GB of HBM2e memory, for a total of 128 GB per GPU. The latter architecture yields eight GPU compute units per node, enabling highly parallel and memory-intensive workloads.

In all the benchmarks executed on the pre-exascale Leonardo HPC, the source code was compiled using the Nvidia Compiler, version 24.3, in conjunction with OpenMPI 4.1.6, an open-source implementation of the Message Passing Interface (MPI) standard. Instead, for the benchmark runs on the LUMI pre-exascale HPC, we used the HPE Cray Fortran Compiler, version 17.0.1, along with the Cray-provided MPI implementation based on MPICH, version 8.1.29.

\begin{figure}[ht]
    \centering
    \includegraphics[width=0.7\linewidth]{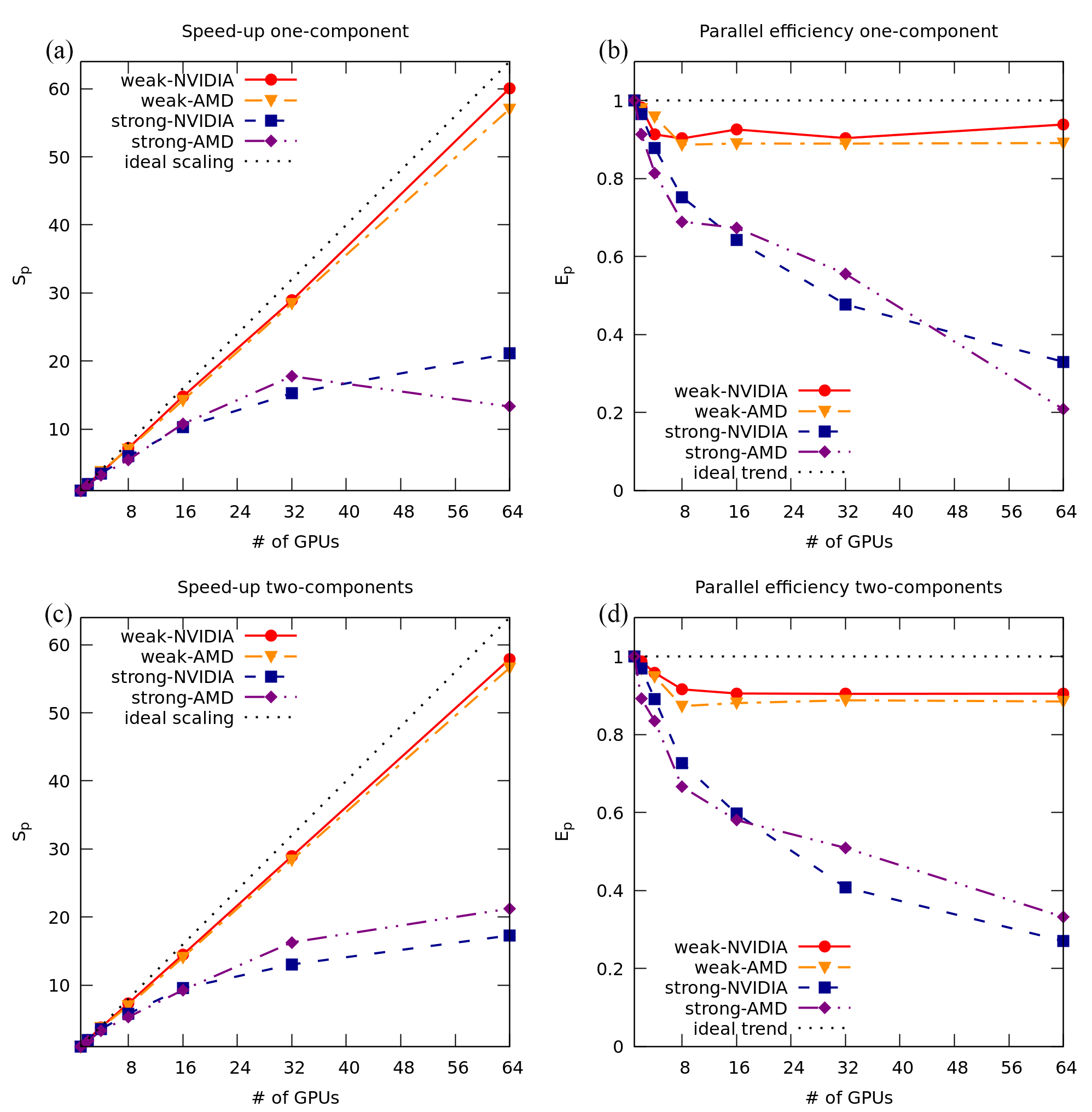}
    \caption{Panels (a) and (b) show the speed up $S_p$ and parallel efficiency $E_p$ for the single component case measured in strong and weak scaling on the NVIDIA Ampere\texttrademark~A100 and AMD Instinct\texttrademark~MI250 GPU-based HPC infrastructures. Panels (c) and (d) report the corresponding data for the two-component benchmark case.}
    \label{fig:benchmark}
\end{figure}

Strong scaling results are presented in Figure \ref{fig:benchmark} in terms of the speedup, $S_{p}$ and parallel efficiency, $E_{p}$, for the single- and double-component systems, evaluated with the MPI decomposition that yields the best performance in terms of GLUPS over the two different GPU-based HPC infrastructures. Note that each AMD Instinct\texttrademark~MI250X card provides two GPU units, and the performance estimators are evaluated as a function of the number of GPU units. The benchmark data indicate that the impact of parallel communication increases as the computational sub-domain of each GPU device diminishes.

To better illustrate the effect of communication overhead, we also report weak scaling results, where the sub-domain size per MPI process (i.e., per GPU device) remains constant. In Figure \ref{fig:benchmark}, we show weak scaling performance using a fixed cubic sub-domain of side 512 lattice points per GPU. This scaling study employs a linear decomposition along the z-axis and demonstrates excellent parallel efficiency, obtaining a peak performance value of 154.94 GLUPS for single-component and 105.48 GLUPS for two-component systems on 64 GPU devices in weak-scaling on NVIDIA Ampere\texttrademark~A100 GPU-based cluster, versus 92.63 GLUPS and 65.67 GLUPS for the single- and double-component systems, respectively, for the AMD Instinct\texttrademark~MI250 GPU-based cluster.

\section{Results}

\subsection{Single-Component Homogeneous Isotropic Turbulence: Energy and Pressure Spectra and Single Point Statistics}

We evaluate the accuracy of the \textit{accLB} code \cite{montessori2024high,lauricella2025thread} in performing fully developed turbulent flow simulations by analyzing single-component homogeneous isotropic turbulence (HIT), focusing on energy and pressure spectra along with selected single-point statistics. We simulate three separate cases of increasing Taylor-scale Reynolds numbers, \(Re_\lambda = 150\), \(230\) and \(370\), on a \(512^3\) uniform Cartesian grid with unit spacing, \(\Delta x = 1\).

The case at \(Re_\lambda = 150\) can be effectively considered as a Direct Numerical Simulation (DNS), where the Kolmogorov length scale is approximately equal to the grid spacing, \(\Delta x\), ensuring adequate resolution of the smallest turbulent scales. Conversely, the simulations at \(Re_\lambda = 230\) and \(370\) lie beyond the strict DNS regime, with a Kolmogorov scale of approximately \(0.4\Delta x\) and \(0.15\Delta x\), respectively.

The homogeneous isotropic turbulent state in all three cases is maintained through large-scale energy injection via the ABC (Arnold–Beltrami–Childress) forcing method. This deterministic, divergence-free forcing scheme introduces a time-independent velocity field that acts as a steady source of energy at the largest scales. Specifically, the forcing takes the form:
\[
\mathbf{f}(\mathbf{x}) = A
\begin{pmatrix}
\sin(k_f z) + \cos(k_f y) \\
\sin(k_f x) + \cos(k_f z) \\
\sin(k_f y) + \cos(k_f x)
\end{pmatrix},
\]
where \(A\) is the forcing amplitude and \(k_f\) the forcing wavenumber. The ABC forcing guarantees a sustained turbulent regime by balancing viscous dissipation with energy input, without introducing significant anisotropy, thereby providing an efficient and widely adopted method to achieve statistically stationary HIT in periodic domains \cite{PODVIGINA1994471}.

\begin{figure}
    \centering
    \includegraphics[width=0.8\linewidth]{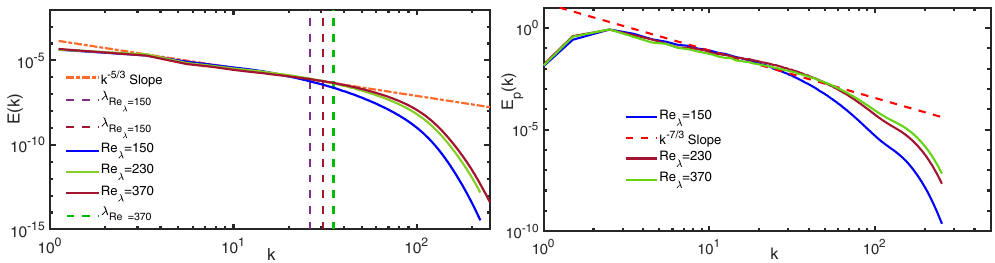}
    \caption{Panel (a) shows the energy spectra for  \(Re_\lambda\), 150, 230 and 370. The expected \(-5/3\) power-law in the inertial range is well captured across a broad range of wavenumbers. Panel (b) displays the corresponding pressure spectra, which follow the \(-7/3\) scaling in the inertial range.}
    \label{fig:spectraSC}
\end{figure}

\begin{figure}
    \centering
    \includegraphics[width=0.8\linewidth]{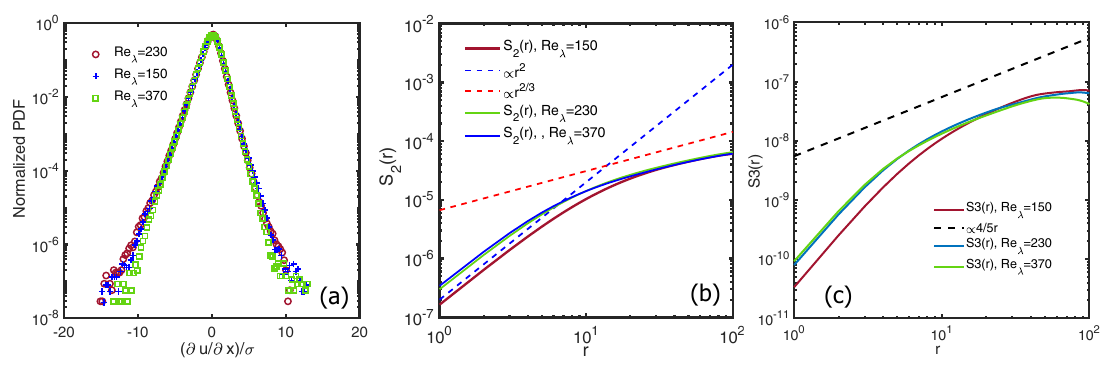}
    \caption{Panel (a): Normalized probability density function (PDF) of the longitudinal velocity gradients. Panel (b): Second-order structure function, compared with the theoretical \(r^2\) scaling in the dissipation range and \(r^{2/3}\) Kolmogorov scaling in the inertial range. Panel (c): Third-order structure function, compared with the theoretical \(r^2\) scaling in the dissipation range and \(r^{2/3}\) Kolmogorov scaling in the inertial range.}
    \label{fig:spsSC}
\end{figure}

Figure~\ref{fig:spectraSC}(a) presents the energy spectra for the cases under scrutiny, compared to the classical \(-5/3\) slope predicted for the inertial subrange \cite{Pope_2000, frisch1995turbulence}. The observed agreement serves as initial validation of the thread-safe LB solver’s capability to reproduce energy transfer mechanisms in single-component turbulent flows.

In figure~\ref{fig:spectraSC}(b), the pressure spectra are plotted and compared to the theoretical \(-7/3\) scaling law \cite{zhao2016}. In both panels, the simulation at higher Reynolds number exhibits a broader spectrum, indicative of an extended inertial range and enhanced scale separation. As expected, deviations from the \(-5/3\) and \(-7/3\) power laws emerge at large scales, near the Taylor length scale, which is marked by vertical dashed lines in the plots.

Figure~\ref{fig:spsSC} shows single-point statistical features of the homogeneous isotropic turbulent field under examination. Panel (a) displays the probability density function (PDF) of the longitudinal velocity gradient at increasing Taylor Reynolds numbers.

As shown in panel (a), the PDFs deviate significantly from a Gaussian distribution and show negative skewness, \(S \approx 0.42\) \((S = \langle (\partial u/\partial x)^3 \rangle / \langle (\partial u/\partial x)^2 \rangle^{3/2})\), and a flatness coefficient \(F \approx 5.25\) \((F = \langle (\partial u/\partial x)^4 \rangle / \langle (\partial u/\partial x)^2 \rangle^{2})\), both in line with prior studies \cite{ishihara2009}. In the physics of HIT, negative skewness points to the presence of long-tails in the statistical distribution of the velocity increments, highlighting the presence of anisotropic effects at small scales, typically caused by persistent vortex stretching, while the elevated flatness signals the presence of intermittent rare-events, i.e., intense fluctuations in the velocity gradients \cite{mccomb1995theory}.

Panel (b) reports the second-order structure function, \(S_2(r) = \langle [u(x_i + r_i) - u(x_i)]^2 \rangle\), benchmarked against two theoretical power laws \cite{mccomb1995theory}: a \(r^2\) scaling at small \(r\), typical of the dissipation range, and the Kolmogorov \(r^{2/3}\) scaling in the inertial range, representative of energy transfer across larger scales. Panel (c) reports the value of the third-order invariant \(S_3(r)\) across scales. In classical Kolmogorov theory, \(S_3(r)\) follows a linear scaling in the inertial range (i.e., \(S_3(r) = -\frac{4}{5} \varepsilon r\), where \(\varepsilon\) is the energy dissipation rate), implying that the energy transfer rate is scale invariant within the inertial range, i.e., that the turbulence exhibits a form of self-similarity across scales \cite{Pope_2000, mccomb1995theory,briscolini1994extended}. Both statistics show agreement with the expected theoretical behavior and tend to deviate from the theoretical behavior as \(r\) increases to values close to the integral length scale \cite{ishihara2009}.

\subsection{Bubble-Laden Homogeneous Isotropic Turbulence: Effect of the dispersed phase on turbulence modulation}

\begin{figure}
    \centering
    \includegraphics[width=0.6\linewidth]{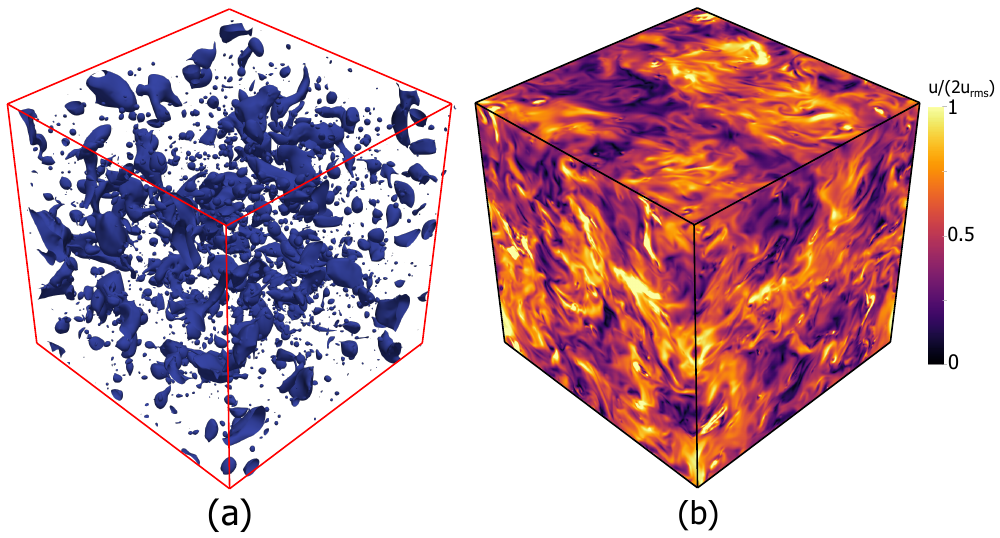}
    \caption{Instantaneous snapshots of the bubble field panel (a) and velocity field (b) at $Re_\lambda=114$ and $6\%$ bubble volume fraction. Snapshots are taken at the statistical steady state. }
    \label{fig:bubbles}
\end{figure}

In this section, we investigate the effect of a dispersed, buoyant gas phase on the turbulence. Specifically, we examine how bubbles influence the energy spectrum at selected wavelengths and explore this behavior across a range of Taylor Reynolds numbers. The bubble volume fraction is fixed at $6\%$ of the total domain, which remains $512^3$, as in the single-phase case (see figure \ref{fig:bubbles}, reporting two snapshots of the bubble and velocity field taken at the statistical steady state). The computed values of $Re_\lambda$ for the cases under consideration are $65$, $86$ and $114$. The surface tension coefficient is set to $\gamma = 0.01$ (in lattice units) and is kept constant across all simulations. The turbulent Weber number, defined as $We_T = u_{rms}^2 \lambda \rho_L / \gamma$ is of order one in all simulations (in the formula $u_{rms}$ is the root mean square value of the velocity fluctuations, $\lambda$ the Taylor length scale and $\rho_L$ the bulk density). The turbulent Froude number, $Fr_T = u_{rms} / \sqrt{g L_I}$, where $L_I$ is the integral length scale, is approximately $2$, indicating that while bubbles are influenced by turbulence, their buoyancy remains sufficiently strong to impose a preferential upward motion.

\begin{figure}
    \centering
    \includegraphics[width=0.55\linewidth]{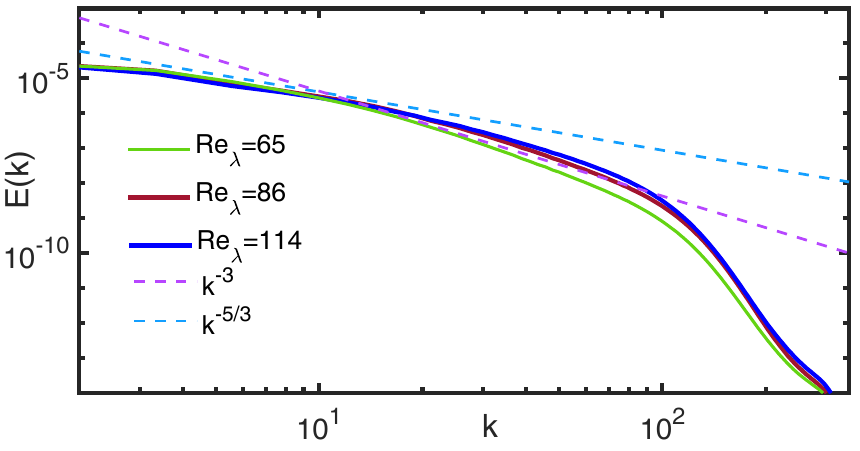}
    \caption{Energy spectra for bubble-laden HIT at different $Re_\lambda$. The transition from $-5/3$ to $-3$ scaling, observed experimentally in~\cite{mercado2010bubble}, is captured in the intermediate wavenumber range.}
    \label{fig:bubble_spectra}
\end{figure}

As shown in Figure~\ref{fig:bubble_spectra}, the energy spectra reveal a clear transition in scaling behavior with increasing $Re_\lambda$, from the classical $-5/3$ Kolmogorov scaling to a steeper $-3$ power law,  typical of 2D turbulence\cite{benzi1990two}. This trend aligns with experimental observations~\cite{mercado2010bubble} and theoretical predictions~\cite{risso2011theoretical}. The emergence of the steeper spectrum is a distinct signature of bubble-induced dissipation at intermediate wavenumbers. Notably, this effect is only observed in buoyant bubble–liquid systems, suggesting that buoyancy plays a critical role in enabling new pathways of energy transfer within homogeneous turbulence. The transition occurs around $k = 10$ (corresponding to $r \approx 35$), approximately one-third of the integral length scale.

\begin{figure}
    \centering
    \includegraphics[width=0.5\linewidth]{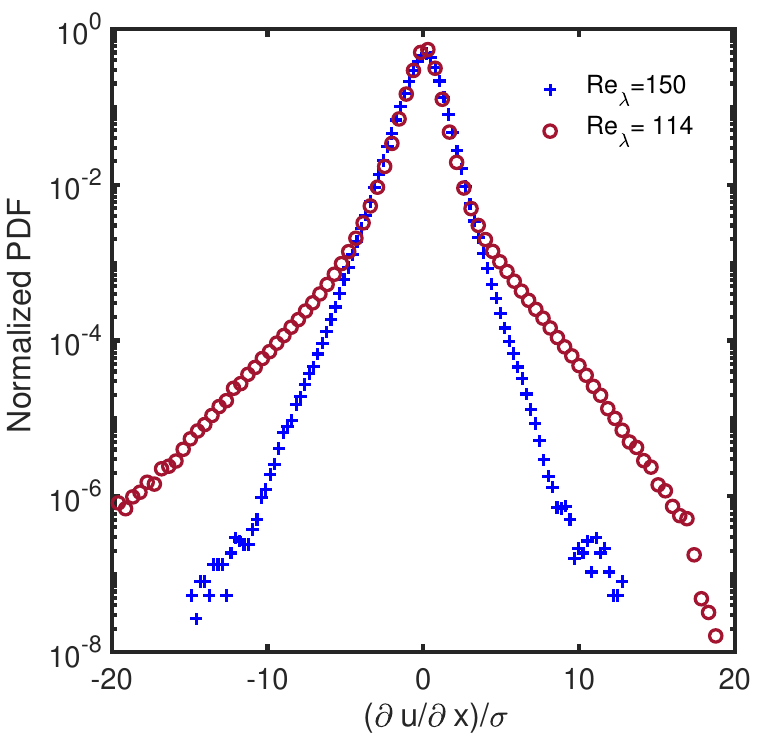}
    \caption{Comparison of velocity increment PDFs between single-phase and bubble-laden HIT. The enhanced tails in the latter case indicate increased intermittency induced by bubble dynamics. In the figure, circles stand for the bubble-laden HIT while plusses for the single component case}
    \label{fig:pdfscomp}
\end{figure}

Risso~\cite{risso2011theoretical} analytically demonstrated that the $-3$ scaling in dispersed multiphase flows arises due to the presence of strong, statistically independent, and uniformly distributed bursts within the turbulent field.
These bursts can be interpreted as intermittent events driven by the complex dynamics of bubbles interacting with the carrier phase. As bubbles deform, break up, and re-coalesce, they continuously reorganize the surrounding flow, thus altering the statistical properties of HIT. This behavior is clearly illustrated in Figure~\ref{fig:pdfscomp}, which compares the probability density functions (PDFs) of velocity increments in the single-phase and bubble-laden cases. The simulation parameters are otherwise identical, though the resulting Taylor Reynolds numbers differ slightly due to enhanced activity at small scales caused by the dispersed phase \cite{crialesi2023interaction}.

While both PDFs display the expected non-Gaussian features of homogeneous isotropic turbulence, the presence of bubbles significantly amplifies the tails of the distribution. This broadening indicates a higher frequency of intense velocity fluctuations, a hallmark of intermittency. The more pronounced tails in the bubble-laden case point to localized, energetic bursts and rare events arising from bubble dynamics. These findings underscore how dispersed bubbles modify the fine-scale structure of turbulence, intensifying gradients and increasing deviations from the Gaussian statistics typically seen in single-phase flows. The presence of strong intermittent events is further supported by the increase of the flatness factor by a factor 3, from approximately $5$ to $16$.

\section{Conclusions}

In this work we have presented accLB, a high-performance, thread-safe Lattice Boltzmann solver specifically tailored for the simulation of multiphase turbulent flows on modern GPU-based supercomputing architectures. By coupling a conservative phase-field formulation of the Allen–Cahn equation with a regularized LBM, accLB have shown to accurately capture the dynamics of fluid interfaces in the presence of turbulence and large density and viscosity contrasts.

The code architecture exploits hybrid parallelism via MPI and OpenACC, offering excellent portability and scalability across heterogeneous hardware platforms. Benchmark tests performed on the EuroHPC pre-exascale systems Leonardo and LUMI demonstrate strong and weak scaling performance up to 64 parallel GPUs, with sustained throughput exceeding 150 GLUPS in large-scale runs.

The physical capabilities of accLB are validated through simulations of single-phase and bubble-laden homogeneous isotropic turbulence (HIT). In the single-phase case, we recover the classical energy and pressure spectra scalings and turbulence invariants across scales. For the bubble-laden scenarios, our simulations capture the emergence of a -3 scaling in the energy spectrum, as predicted by theory and observed in experiments, along with amplified small-scale intermittency. These findings confirm that accLB is a reliable tool to investigate turbulence modulation in complex multiphase systems.

In summary, accLB provides a scalable and accurate platform for high-fidelity turbulence simulations in multiphase flows, enabling the investigation of complex physical phenomena across a broad range of scales with high computational efficiency.

\section*{Acknowledgements}

A.M. acknowledges fundings from the Italian Government through the PRIN (Progetti di Rilevante Interesse Nazionale) Grant (MOBIOS) ID: 2022N4ZNH3 -CUP: F53C24001000006 and computational support of CINECA through the ISCRA B project DODECA (HP10BUBFIL). M.L. acknowledges the support of the Italian National Group for Mathematical Physics (GNFM-INdAM). M.L. and S.S. acknowledge the support from the European Research Council under the ERCPoC Grant No. 101187935 (LBFAST). D.I. acknowledges financial support from the Research Council of Finland (Grant number: 354620). We acknowledge CSC, Finland for awarding this project access to the LUMI supercomputer, owned by the EuroHPC Joint Undertaking, hosted by CSC (Finland) and the LUMI consortium through CSC, Finland, Benchmark Access mode.

\bibliographystyle{plainnat}
\bibliography{mybibfile}

\begin{thebibliography}{43}
\providecommand{\natexlab}[1]{#1}
\providecommand{\url}[1]{\texttt{#1}}
\expandafter\ifx\csname urlstyle\endcsname\relax
  \providecommand{\doi}[1]{doi: #1}\else
  \providecommand{\doi}{doi: \begingroup \urlstyle{rm}\Url}\fi

\bibitem[Benzi and Succi(1990)]{benzi1990two}
R~Benzi and S~Succi.
\newblock Two-dimensional turbulence with the lattice boltzmann equation.
\newblock \emph{Journal of Physics A: Mathematical and General}, 23\penalty0 (1):\penalty0 L1, 1990.

\bibitem[Bernardini et~al.(2021)Bernardini, Modesti, Salvadore, and Pirozzoli]{Bernardini2021}
Matteo Bernardini, Davide Modesti, Francesco Salvadore, and Sergio Pirozzoli.
\newblock Streams: A high-fidelity accelerated solver for direct numerical simulation of compressible turbulent flows.
\newblock \emph{Computer Physics Communications}, 263:\penalty0 107906, 2021.
\newblock ISSN 0010-4655.

\bibitem[Bonaccorso et~al.(2022)Bonaccorso, Lauricella, Montessori, Amati, Bernaschi, Spiga, Tiribocchi, and Succi]{Bonaccorso2022}
Fabio Bonaccorso, Marco Lauricella, Andrea Montessori, Giorgio Amati, Massimo Bernaschi, Filippo Spiga, Adriano Tiribocchi, and Sauro Succi.
\newblock Lbcuda: A high-performance cuda port of lbsoft for simulation of colloidal systems.
\newblock \emph{Computer Physics Communications}, 277:\penalty0 108380, 2022.
\newblock ISSN 0010-4655.

\bibitem[Brandt and Coletti(2022)]{brandt2022ARFM}
Luca Brandt and Filippo Coletti.
\newblock Particle-laden turbulence: Progress and perspectives.
\newblock \emph{Annual Review of Fluid Mechanics}, 54\penalty0 (Volume 54, 2022):\penalty0 159--189, 2022.
\newblock ISSN 1545-4479.

\bibitem[Briscolini et~al.(1994)Briscolini, Santangelo, Succi, and Benzi]{briscolini1994extended}
M~Briscolini, P~Santangelo, S~Succi, and R~Benzi.
\newblock Extended self-similarity in the numerical simulation of three-dimensional homogeneous flows.
\newblock \emph{Physical Review E}, 50\penalty0 (3):\penalty0 R1745, 1994.

\bibitem[Costa et~al.(2021)Costa, Phillips, Brandt, and Fatica]{costa2021}
Pedro Costa, Everett Phillips, Luca Brandt, and Massimiliano Fatica.
\newblock Gpu acceleration of cans for massively-parallel direct numerical simulations of canonical fluid flows.
\newblock \emph{Computers and Mathematics with Applications}, 81:\penalty0 502--511, 2021.
\newblock ISSN 0898-1221.
\newblock Development and Application of Open-source Software for Problems with Numerical PDEs.

\bibitem[Crialesi-Esposito et~al.(2023{\natexlab{a}})Crialesi-Esposito, Chibbaro, and Brandt]{crialesi2023interaction}
Marco Crialesi-Esposito, Sergio Chibbaro, and Luca Brandt.
\newblock The interaction of droplet dynamics and turbulence cascade.
\newblock \emph{Communications Physics}, 6\penalty0 (1):\penalty0 5, 2023{\natexlab{a}}.

\bibitem[Crialesi-Esposito et~al.(2023{\natexlab{b}})Crialesi-Esposito, Scapin, Demou, Rosti, Costa, Spiga, and Brandt]{crialesi2023flutas}
Marco Crialesi-Esposito, Nicolò Scapin, Andreas~D. Demou, Marco~Edoardo Rosti, Pedro Costa, Filippo Spiga, and Luca Brandt.
\newblock Flutas: A gpu-accelerated finite difference code for multiphase flows.
\newblock \emph{Computer Physics Communications}, 284:\penalty0 108602, 2023{\natexlab{b}}.
\newblock ISSN 0010-4655.

\bibitem[{De Vanna}(2025)]{devanna2025}
Francesco {De Vanna}.
\newblock Enhancing performance of high-speed engineering flow computations: The uranos case study.
\newblock \emph{Procedia Computer Science}, 255:\penalty0 23--32, 2025.
\newblock ISSN 1877-0509.
\newblock Proceedings of the Second EuroHPC user day.

\bibitem[Fakhari et~al.(2017)Fakhari, Mitchell, Leonardi, and Bolster]{fakhari2017improved}
Abbas Fakhari, Travis Mitchell, Christopher Leonardi, and Diogo Bolster.
\newblock Improved locality of the phase-field lattice-boltzmann model for immiscible fluids at high density ratios.
\newblock \emph{Physical Review E}, 96\penalty0 (5):\penalty0 053301, 2017.

\bibitem[Feng et~al.(2019{\natexlab{a}})Feng, Boivin, Jacob, and Sagaut]{feng2019hybrid}
Yongliang Feng, Pierre Boivin, J{\'e}r{\^o}me Jacob, and Pierre Sagaut.
\newblock Hybrid recursive regularized lattice boltzmann simulation of humid air with application to meteorological flows.
\newblock \emph{Physical Review E}, 100\penalty0 (2):\penalty0 023304, 2019{\natexlab{a}}.

\bibitem[Feng et~al.(2019{\natexlab{b}})Feng, Boivin, Jacob, and Sagaut]{feng2019hybrid2}
Yongliang Feng, Pierre Boivin, J{\'e}r{\^o}me Jacob, and Pierre Sagaut.
\newblock Hybrid recursive regularized thermal lattice boltzmann model for high subsonic compressible flows.
\newblock \emph{Journal of Computational Physics}, 394:\penalty0 82--99, 2019{\natexlab{b}}.

\bibitem[Frisch and Kolmogorov(1995)]{frisch1995turbulence}
Uriel Frisch and Andrei~Nikolaevich Kolmogorov.
\newblock \emph{Turbulence: the legacy of AN Kolmogorov}.
\newblock Cambridge university press, 1995.

\bibitem[Grabowski and Wang(2013)]{grabowski2013growth}
Wojciech~W Grabowski and Lian-Ping Wang.
\newblock Growth of cloud droplets in a turbulent environment.
\newblock \emph{Annual review of fluid mechanics}, 45\penalty0 (1):\penalty0 293--324, 2013.

\bibitem[Gropp(1992)]{gropp1992parallel}
William~D Gropp.
\newblock Parallel computing and domain decomposition.
\newblock In \emph{Fifth International Symposium on Domain Decomposition Methods for Partial Differential Equations}, pages 349--361. Publ by Soc for Industrial and Applied Mathematics Publ, 1992.

\bibitem[Guo et~al.(2002)Guo, Zheng, and Shi]{guo2002discrete}
Zhaoli Guo, Chuguang Zheng, and Baochang Shi.
\newblock Discrete lattice effects on the forcing term in the lattice boltzmann method.
\newblock \emph{Physical review E}, 65\penalty0 (4):\penalty0 046308, 2002.

\bibitem[Hennessy and Patterson(2011)]{hennessy2011computer}
John~L Hennessy and David~A Patterson.
\newblock \emph{Computer architecture: a quantitative approach}.
\newblock Elsevier, 2011.

\bibitem[Ishihara et~al.(2009)Ishihara, Gotoh, and Kaneda]{ishihara2009}
Takashi Ishihara, Toshiyuki Gotoh, and Yukio Kaneda.
\newblock Study of high–reynolds number isotropic turbulence by direct numerical simulation.
\newblock \emph{Annual Review of Fluid Mechanics}, 41:\penalty0 165--180, 2009.
\newblock \doi{10.1146/annurev.fluid.010908.165203}.

\bibitem[Jiang and Shu(1996)]{JIANG1996202}
Guang-Shan Jiang and Chi-Wang Shu.
\newblock Efficient implementation of weighted eno schemes.
\newblock \emph{Journal of Computational Physics}, 126\penalty0 (1):\penalty0 202--228, 1996.
\newblock ISSN 0021-9991.
\newblock \doi{https://doi.org/10.1006/jcph.1996.0130}.
\newblock URL \url{https://www.sciencedirect.com/science/article/pii/S0021999196901308}.

\bibitem[Jähne and Haußecker(1998)]{jahne1998}
B.~Jähne and H.~Haußecker.
\newblock Air-water gas exchange.
\newblock \emph{Annual Review of Fluid Mechanics}, 30\penalty0 (Volume 30, 1998):\penalty0 443--468, 1998.
\newblock ISSN 1545-4479.

\bibitem[Kim(2005)]{kim2005continuous}
Junseok Kim.
\newblock A continuous surface tension force formulation for diffuse-interface models.
\newblock \emph{Journal of computational physics}, 204\penalty0 (2):\penalty0 784--804, 2005.

\bibitem[Kr{\"u}ger et~al.(2017)Kr{\"u}ger, Kusumaatmaja, Kuzmin, Shardt, Silva, and Viggen]{kruger2017lattice}
Timm Kr{\"u}ger, Halim Kusumaatmaja, Alexandr Kuzmin, Orest Shardt, Goncalo Silva, and Erlend~Magnus Viggen.
\newblock The lattice boltzmann method.
\newblock \emph{Springer International Publishing}, 10\penalty0 (978-3):\penalty0 4--15, 2017.

\bibitem[Lauricella et~al.(2025)Lauricella, Tiribocchi, Succi, Brandt, Mukherjee, La~Rocca, and Montessori]{lauricella2025thread}
Marco Lauricella, Adriano Tiribocchi, Sauro Succi, Luca Brandt, Aritra Mukherjee, Michele La~Rocca, and Andrea Montessori.
\newblock Thread-safe multiphase lattice boltzmann model for droplet and bubble dynamics at high density and viscosity contrasts.
\newblock \emph{arXiv preprint arXiv:2501.00846}, 2025.

\bibitem[Lupo et~al.(2025)Lupo, Wellens, and Costa]{lupo2025}
Giandomenico Lupo, Peter Wellens, and Pedro Costa.
\newblock Cans-fizzy: A gpu-accelerated finite difference solver for turbulent two-phase flows.
\newblock \emph{arXiv preprint arXiv:2502.04189}, 2025.

\bibitem[Malaspinas(2015)]{malaspinas2015increasing}
Orestis Malaspinas.
\newblock Increasing stability and accuracy of the lattice boltzmann scheme: recursivity and regularization.
\newblock \emph{arXiv preprint arXiv:1505.06900}, 2015.

\bibitem[McComb(1995)]{mccomb1995theory}
WD~McComb.
\newblock Theory of turbulence.
\newblock \emph{Reports on Progress in Physics}, 58\penalty0 (10):\penalty0 1117, 1995.

\bibitem[Mercado et~al.(2010)Mercado, Gomez, Van~Gils, Sun, and Lohse]{mercado2010bubble}
Julian~Martinez Mercado, Daniel~Chehata Gomez, Dennis Van~Gils, Chao Sun, and Detlef Lohse.
\newblock On bubble clustering and energy spectra in pseudo-turbulence.
\newblock \emph{Journal of fluid mechanics}, 650:\penalty0 287--306, 2010.

\bibitem[Montessori and Falcucci(2018)]{montessori2018lattice}
Andrea Montessori and Giacomo Falcucci.
\newblock \emph{Lattice Boltzmann modeling of complex flows for engineering applications}.
\newblock Morgan and Claypool Publishers, 2018.

\bibitem[Montessori et~al.(2023)Montessori, Lauricella, Tiribocchi, Durve, La~Rocca, Amati, Bonaccorso, and Succi]{montessori2023thread}
Andrea Montessori, Marco Lauricella, Adriano Tiribocchi, Mihir Durve, Michele La~Rocca, Giorgio Amati, Fabio Bonaccorso, and Sauro Succi.
\newblock Thread-safe lattice boltzmann for high-performance computing on gpus.
\newblock \emph{Journal of Computational Science}, 74:\penalty0 102165, 2023.

\bibitem[Montessori et~al.(2024)Montessori, La~Rocca, Amati, Lauricella, Tiribocchi, and Succi]{montessori2024high}
Andrea Montessori, Michele La~Rocca, Giorgio Amati, Marco Lauricella, Adriano Tiribocchi, and Sauro Succi.
\newblock High-order thread-safe lattice boltzmann model for high performance computing turbulent flow simulations.
\newblock \emph{Physics of Fluids}, 36\penalty0 (3), 2024.

\bibitem[Montessori et~al.(2025)Montessori, Hegele, and Lauricella]{montessori2025thread}
Andrea Montessori, Luiz~A Hegele, and Marco Lauricella.
\newblock A thread-safe lattice boltzmann model for multicomponent turbulent jet simulations.
\newblock \emph{AIAA Journal}, 63\penalty0 (3):\penalty0 1005--1012, 2025.

\bibitem[Pelusi et~al.(2022)Pelusi, Lulli, Sbragaglia, and Bernaschi]{pelusi2022tlbfind}
Francesca Pelusi, Matteo Lulli, Mauro Sbragaglia, and Massimo Bernaschi.
\newblock Tlbfind: a thermal lattice boltzmann code for concentrated emulsions with finite-size droplets.
\newblock \emph{Computer Physics Communications}, 273:\penalty0 108259, 2022.

\bibitem[Pelusi et~al.(2023)Pelusi, Ascione, Sbragaglia, and Bernaschi]{pelusi2023analysis}
Francesca Pelusi, Stefano Ascione, Mauro Sbragaglia, and Massimo Bernaschi.
\newblock Analysis of the heat transfer fluctuations in the rayleigh--b{\'e}nard convection of concentrated emulsions with finite-size droplets.
\newblock \emph{Soft Matter}, 19\penalty0 (37):\penalty0 7192--7201, 2023.

\bibitem[Podvigina and Pouquet(1994)]{PODVIGINA1994471}
O.~Podvigina and A.~Pouquet.
\newblock On the non-linear stability of the 1:1:1 abc flow.
\newblock \emph{Physica D: Nonlinear Phenomena}, 75\penalty0 (4):\penalty0 471--508, 1994.
\newblock ISSN 0167-2789.
\newblock \doi{https://doi.org/10.1016/0167-2789(94)00031-X}.
\newblock URL \url{https://www.sciencedirect.com/science/article/pii/016727899400031X}.

\bibitem[Pope(2000)]{Pope_2000}
Stephen~B. Pope.
\newblock \emph{Turbulent Flows}.
\newblock Cambridge University Press, 2000.

\bibitem[Risso(2011)]{risso2011theoretical}
Fr{\'e}d{\'e}ric Risso.
\newblock Theoretical model for k- 3 spectra in dispersed multiphase flows.
\newblock \emph{Physics of fluids}, 23\penalty0 (1), 2011.

\bibitem[Roccon(2024)]{Roccon2024}
Alessio Roccon.
\newblock A gpu-ready pseudo-spectral method for direct numerical simulations of multiphase turbulence.
\newblock \emph{Procedia Computer Science}, 240:\penalty0 17--30, 2024.
\newblock ISSN 1877-0509.
\newblock Proceedings of the First EuroHPC user day.

\bibitem[Sardina et~al.(2015)Sardina, Picano, Brandt, and Caballero]{sardina2015}
Gaetano Sardina, Francesco Picano, Luca Brandt, and Rodrigo Caballero.
\newblock Continuous growth of droplet size variance due to condensation in turbulent clouds.
\newblock \emph{Phys. Rev. Lett.}, 115:\penalty0 184501, Oct 2015.

\bibitem[Schramm et~al.(2003)Schramm, Stasiuk, and Marangoni]{schramm2003}
Laurier~L. Schramm, Elaine~N. Stasiuk, and D.~Gerrard Marangoni.
\newblock Surfactants and their applications.
\newblock \emph{Annu. Rep. Prog. Chem.{,} Sect. C: Phys. Chem.}, 99:\penalty0 3--48, 2003.

\bibitem[Succi(2018)]{succi}
S.~Succi.
\newblock \emph{The Lattice Boltzmann equation: For complex states of flowing matter}.
\newblock Oxford University Press, 2018.

\bibitem[Succi et~al.(2019)Succi, Amati, Bernaschi, Falcucci, Lauricella, and Montessori]{exasc2}
S.~Succi, G.~Amati, M.~Bernaschi, G.~Falcucci, M.~Lauricella, and A.~Montessori.
\newblock Towards exascale lattice boltzmann computing.
\newblock \emph{Computer and Fluids}, 181:\penalty0 107--115, 2019.
\newblock \doi{https://doi.org/10.1016/j.compfluid.2019.01.005}.

\bibitem[Zhao et~al.(2016)Zhao, Cheng, Qiu, Burnett, and chun Liu]{zhao2016}
Sipei Zhao, Eva Cheng, Xiaojun Qiu, Ian Burnett, and Jacob~Chia chun Liu.
\newblock Pressure spectra in turbulent flows in the inertial and the dissipation ranges.
\newblock \emph{The Journal of the Acoustical Society of America}, 140:\penalty0 4178, 2016.
\newblock \doi{10.1121/1.4968881}.

\bibitem[Zhu et~al.(2018)Zhu, Phillips, Spandan, Donners, Ruetsch, Romero, Ostilla-Mónico, Yang, Lohse, Verzicco, Fatica, and Stevens]{Zhu_AFiD_2018}
Xiaojue Zhu, Everett Phillips, Vamsi Spandan, John Donners, Gregory Ruetsch, Joshua Romero, Rodolfo Ostilla-Mónico, Yantao Yang, Detlef Lohse, Roberto Verzicco, Massimiliano Fatica, and Richard~J.A.M. Stevens.
\newblock Afid-gpu: A versatile navier–stokes solver for wall-bounded turbulent flows on gpu clusters.
\newblock \emph{Computer Physics Communications}, 229:\penalty0 199--210, 2018.
\newblock ISSN 0010-4655.

\end{thebibliography}

\end{document}